\documentclass[12pt]{article}
\usepackage{amssymb}
\usepackage{epsfig,color}
\usepackage{pstricks,graphicx,epsfig,color,amssymb,amsmath,amscd}
\usepackage{cite}
\usepackage[]{graphicx}
\usepackage{makeidx}
\usepackage{amsmath}
\usepackage{nccmath}
\usepackage{setspace}
\usepackage{ulem}

\usepackage[]{caption}  
\renewcommand\rho{\varrho}
\newcommand{\bi}{\bibitem}
\newcommand{\be}{\begin{eqnarray}}
\newcommand{\ee}{\end{eqnarray}}
\newcommand{\rar}{\rightarrow}

\newcommand{\tcblue}{\textcolor{blue}}

\newcommand{\tcmag}{\textcolor{magenta}}
\newcommand{\tcred}{\textcolor{red}}
\newcommand{\tcviol}{\textcolor{violet}}
\newcommand{\tccyan}{\textcolor{cyan}}

\begin{document}
\begin{titlepage}
\title{
James Webb Space Telescope: data, problems, and resolution
{
}\\
\small{
Plenary talk at 36th Rencontres de Physique de la Vall{\'e}e d'Aoste \\
on Results and Perspectives in Particle Physics,\\
extended and updated 
}
}
\author{ A.\,D. Dolgov$^{a,b}$}

\maketitle
\begin{center}
$^a${Department of Physics, Novosibirsk State University,\\ 
Pirogova st.\,2, Novosibirsk, 630090 Russia}\\
$^b${{Bogolyubov Laboratory of Theoretical Physics, Joint Institute for Nuclear Research,\\
Joliot-Curie st.\,6, Dubna, Moscow region, 141980 Russia}}\\
\end{center}

\begin{abstract}

It is argued that the data presented  by Hubble Space Telescope and James Webb Space Telescope, that 
seem to be  at odds with the canonical big bang cosmology, find simple explanation if galaxy formation is 
seeded by massive primordial black holes (PBH), as anticipated in 1993 (A. Dolgov and J. Silk, later DS). 
The statement that the galaxy formation might be seeded by PBH is now rediscovered in several  works. 
The predicted by DS log-normal mass spectrum of PBHs very well agrees with astronomical data. Abundant 
BH population of the Galaxy with masses of the order of tens solar masses is predicted. Extended mass 
spectrum of PBH together with their possible clustering allows them to make 100\% contribution into the 
cosmological dark matter. Another prediction of DS mechanism on noticeable amount of antimatter in the 
Milky Way also seems to be confirmed by the data.

\end{abstract}
\thispagestyle{empty}
\end{titlepage}


\section{Introduction \label{s-intro}}

Observations of the last decades by several astronomical instruments, 
especially by the Atacama Large Millimeter/submillimeter Array (ALMA), 
Hubble Space Telescope (HST), and very recently  by James Webb Space 
Telescope (JWST) revealed strong tension between the data and the accepted 
cosmological model. It was discovered that the early universe at redshifts 
$z \sim 10$ and the age of a few hundred million years was densely populated by  
all kinds of astronomical objects: galaxies, quasars (alias supermassive black holes), 
gamma-bursters,  supernovae, and, in addition, it happened to be extremely  dusty.
Equally puzzling problems arose from the  observational data on the present day universe. 

The discussion of the problems, which appeared in the early as well 
as in the contemporary universe, 
at the stage of art, that existed 5 years ago, can be found in review~\cite{AD-UFN}.
But since that time much higher amount of surprising phenomena
was accumulated thanks to new and more precise astronomical instruments.

Possible and, as it seems, simple and natural solution to all these problems has been
suggested, long before they struck the community, by the proposal that the
universe is abundantly populated with primordial black holes. 
A new mechanism of massive PBH creation was worked out
in our paper~\cite{AD-JS}, allowing for formation of highly massive primordial black hole (PBH). 
This mechanism was further elaborated in ref.~\cite{DKK}.
A very simple log-normal mass spectrum of PBH was predicted that, as later verified, very well
agrees with observational data. The abundant PBH population in very wide mass interval can 
eliminate the tension between theory and observations, in particular, because supermassive PBH
could seed galaxy formation as it was envisaged  in refs.~\cite{AD-JS,DKK}.

In addition to high mass PBH formation, the mechanism of refs~\cite{AD-JS,DKK} could lead
to noticeable antimatter population of galaxies. In particular antimatter, including antistars, may exist in
the Milky Way. This exciting prediction seems to be confirmed by the recent studies.\\[2mm]
{\bf Outline of the talk.}\\
1. Recent problems discovered by HST and  JWST. \\
2. Earlier established cosmological problems. \\
3. PBH solution of new and old problems.\\
4. Antimatter in the MIky Way, antistars, anit-nuclej, positrons.\\
5. Log-normal mass spectrum of PBHs. comparison to observations.\\
6. Black dark matter.\\
7. Gravitational waves and PBH.\\
8. {Basics of the mechanism of PBH and antimatter creation .}

\section{Very young galaxies observed by JWST (and HST)  \label{s-blou-by-JWST}}

Discoveries of several recent months made by JWST,  {in continuous  infrared $\mu$m range,}
created almost panic among traditional cosmologists
and astrophysicists. It was observed that the pretty young universe with the age 200-300 
million years  contained a large array of bright galaxies~\cite{JW-1} -\cite{JW-9}, 
which simply { could not be there} according to the accepted
faith or, better to say, to the canonic cosmological model.

{As is stated in the JWST publications:
''an unexpectedly large density (stellar mass density 
$\rho_* \gtrsim 10^6 M_\odot$ Mpc$^{-3}$) of massive galaxies 
(stellar masses $M_* \geq 10^{10.5} M_\odot $)  are discovered at extremely high redshifts $z \gtrsim 10$.''}
Two galaxies with record redshift according to the  Cosmic Evolution Early Release Science (CEERS) data
have redshifts $z = 14.3 \pm 0.4$ that corresponds to the universe age $t_U = 264 $ Myr 
 and $z =16.7$, even younger, at  $t_U =235 $~Myr. 

The early JWST data were taken somewhat sceptically because of lacking direct measurement of the galactic spectra  
that is impossible when working in micron infrared continuum. Measurements of the galactic redshifts by spectral line 
identifications was strongly desirable. An important confirmation of existence of
galaxies in so young universe came from the HST 
{\bf spectroscopic} observation of the most distant galaxy discovered HST at redshift $11.58 \pm 0.05$~\cite{HST-record}.

During the last few weeks there appeared several publications of spectroscopic identification of the far away galaxies.
In ref.~\cite{tacchella} the data from JWST NIRCam 9-band near-infrared imaging of the luminous $z=10.6$
galaxy GN-z11 from the JWST Advanced Deep Extragalactic Survey are presented. 
The authors have concluded that a spectral energy distribution 
is determined entirely consistent with the expected form of the high-redshift galaxy.

At the publication~\cite{bunker} of the same date the spectroscopy of GN-z11, the most luminous
candidate $z>10$ Lyman break galaxy in the GOODS-North field is presented.
Redshift  of $z=10.603$  is derived (somewhat lower than previous
determinations) based on multiple emission lines in low and medium
resolution spectra over $0.8-5.3\,\mu$m. 
The spatially-extended Lyman-$\alpha$ in emission is observed. 
The NIRSpec spectroscopy confirms that GN-z11 is a remarkable galaxy with extreme
properties seen 430 Myr after the Big Bang.

ALMA has confirmed the age of  one of the most distant JWST-identified 
galaxy, GHZ2/GLASS-z12, equal to 367 million years after the Big Bang~\cite{alma-age}. 
by deep spectroscopic 
observations of the spectral emission line associated with {ionized oxygen} near the galaxy.

Spectroscopic observations confirm JWST early galaxy discovery beyond any doubts.

\subsection{Seeding galaxy formation by PBH \label{ss-seeding}}

According to the standard approach the supermassive black holes (SMBHs)
in galactic centres are formed by accretion mechanism
after galaxies were created. In papers.~\cite{AD-JS,DKK} the validity of
the opposite scenario was conjectured, namely, SMBHs were formed first and subsequently seeded
galaxy formation. The hypothesis advocated in these works
allows to explain presence of SMBH in all large and several small galaxies accessible to observation 
and resolves the problem of very early existence of galaxies observed by HST and JWST.

The advocated in this talk and suggested in our earlier papers~\cite{AD-JS,DKK} idea that the galaxy 
are seeded by 
massive black hole seems to gain more and more support in recent publications. Observation of 
supermassive black holes (SMBH) inside galaxies also
indirectly confirms the idea of seeding the galaxy formation by SMBH

This above statement was rediscovered in ref.~\cite{bromm}: 
"...we show that the observed massive galaxy candidates can be explained with lower SFE
than required in $\Lambda\rm CDM$, 
{if structure formation is accelerated by
massive ($\gtrsim 10^{9}\ \rm M_{\odot}$) primordial black holes}
{ that enhance primordial density fluctuations."}

Very recently, in December, 2022, there appeared another paper on the possibility of SMBH impact on
 JWST-galaxy formation~\cite{volontieri}.


According to ref.~\cite{gal-impossible}, six  very well  developed 
galaxies are observed, including one galaxy with a possible stellar mass of $\sim 10^{11} M_\odot$,
at the redshifts $ 7.4 \lesssim z \lesssim 9.1$,  corresponding to 500–700 Myr after the Big Bang. 
These galaxies are too massive
to be created in so early universe. According to the existing science it is impossible to 
create such well developed galaxies 
in this very short time. The authors even make the conclusion:
''May be they are {\bf supermassive black holes of the kind never seen before} 
That might mean a revision of our understanding of black holes.''.
This statement perfectly agrees with the advocated in this talk point of view that massive 
{\bf primordial} black holes populate early universe and seed galaxy formation.

Recently an ultra-massive QSO at $z = 6.853$ was observed by ALMA~\cite{alma-BH}:
"{VIRCam and IRAC photometry perhaps suggests that COS-87259 is an extremely massive reionization-era 
galaxy with $M_* = 1.7 \times 10^{11} M_\odot $}.
Such a very high AGN luminosity suggests that this object is powered by $ \sim 1.6 \times  10^9 M_\odot $ black 
hole if accreting near the Eddington limit." This looks as nearly impossible, but if there is a primordial  
supermassive black hole, it could easily seed such monstrous galaxy and quasar.

In paper~\cite{seed-1}, published already after the Conference was over,
the discovery of an accreting supermassive black hole at z=8.679 is announced
in a galaxy previously obsered via a Ly$\alpha$-break by Hubble
and with a Ly$\alpha$ redshift from Keck. The mass of the black hole is
log($M_{BH}/M_{\odot})=6.95{\pm}0.37$, and it is estimated that it is accreting at
1.2 ($\pm$0.5) times of the Eddington limit. According to the authors:
this presently highest-redshift AGN discovery is used to
place constraints on black hole seeding models and find that a combination of
either super-Eddington accretion from stellar seeds or Eddington accretion from
massive black hole seeds is required to form this object by the observed epoch.

In the paper~\cite{gal-seed} the seeding of galaxies was even indicated in the title of the work.
The authors suggest existence of relatively light PBH with masses about $50 M_\odot$ gaining
mass through super-Eddington accretion within the dark matter halo can explain  observations
of massive galaxies at redshifts of $z\geq 6.5$ by JWST. However, it seems that supermassive PBH formation 
is much less cumbersome. 


 \section{Comment on LSS formation \label{s-LSS}}

According to the canonical theory of large scale structure (LSS) formation the density contrast 
$\Delta \equiv \delta \rho /\rho$ started to rise
at the onset of the matter dominated stage at $z = 10^4$. 
After that $\Delta$ evolved as the cosmological scale factor. Since initially
$\Delta_{in} \lesssim 10^{-4}$, by the present time it may reach unity and after 
that fast LSS formation takes place 
(violent relaxation - strong rising of the gravitational field of the inhomogeneity)
leading to the
observed highly inhomogeneous universe at the galactic and galaxy clusters scales.

In a simple way the process of structure formation
can be understood as follows. The velocity of the Hubble 
runaway at distance $r$ is  $v_H = H r$ and
the virial velocity in the gravitational field of the inhomogeneity is
\be 
v^2_{grav} = \frac{4\pi r^3}{r m_{Pl}^2} \delta \rho
\ee
Using $H^2 = (8 \pi \rho)/(3 m_{pl}^2) $ we find $v_{grav} \geq v_H $ if $\Delta > 1$. 
{ The probability of such huge density fluctuation
for the flat spectrum of perturbations is quite low. 

There are two effects operating in the same directions. 
Firstly, the available time is constrained by the universe age, which is essentially equal to $t_U = 1/H$ 
and is quite short. In addition, a large value of $H$ means expansion is very fast. That strongly 
suppresses the efficiency of the structure formation.}

\section{Problems preceding JWSP \label{s-before-JWST}}
 
 Similar serious problems are known already for several years. 
 The Hubble space telescope discovered that the early universe, at $z = 6-7$ 
 was too densely populated with quasars, alias SMBH, supernovae, gamma-bursters and happened to be
 very dusty. No understanding is found in conventional
 cosmology how all these creature were given birth in such a short time.  
 Moreover  great lots of phenomena in the present day universe, $\sim 15$ billion year old, 
 are also in strong tension with the conventional cosmological expectations.

HST sees the universe up to  $z = 6-7$, but accidentally a galaxy at
$z \approx 12$ has been discovered for which both Hubble and Webb are in good agreement, as we have 
already mentioned in the previous section.
Still, despite the earlier discoveries by HST, only  after publications of JWST data the 
astronomical establishment became seriously worried.

To summarise: observational data of the last decades
present more and more evidence indicating existence of the objects contradicting 
conventional astrophysics and cosmology in the present day and in quite young universe. 
Rephrasing  Marcellus from "The Tragedy of Hamlet, Prince of Denmark" we can say 
"Something is rotten in the state of   \sout{Denmark}  the Universe". 
{However, all the problems can be neatly solved if  the universe is sufficiently dencely
populated by primordial black holes. }

\section{BH types by formation mechanisms \label{s-BH-by-form}}

There three known types of BH depending upon the mechanism of their creation:
\begin{enumerate}
\item{ Astrophysical  black holes.}\\
{These BHs are created by the collapse of a star which exhausted its nuclear fuel.}
The expected masses should start immediately above the neutron star mass, i.e. about
{ ${3M_\odot}$, but noticeably below $100 M_\odot$.}
Instead we observe that the BH mass spectrum in the galaxy has maximum at
  ${M \approx 8 M_\odot}$ with the width  $ \sim(1-2) M_\odot $.   
 The result is somewhat unexpected but an explanations in the conventional astrophysical frameworks is
 not excluded.  \\
  {Recently LIGO/Virgo discovered BHs with masses close to $100 M_\odot$. }
 Their astrophysical origin was considered to be completely
 impossible. Now some, though  quite exotic, formation mechanisms have been suggested. 

\item{Accretion created BHs.}\\
{Such BHs are formed by the accretion of matter on the mass excess in galactic centres.}
It is known that in any large galaxy at the centre there exists a supermassive black holes  (SMBH) with masses
varying from a few millions $M_\odot$ (e,g, Milky Way) up to almost hundred billions $M_\odot$.
However, the conventional accretion mechanisms are not efficient enough to create such
monsters during the universe life-time, ${ t_U \approx 14.6 }$ Gyr. At least 10-fold longer time
is necessary, some references can be found in~\cite{AD-UFN},
{to say nothing about SMBH in 10 times younger universe.}

\item{ Primordial black holes (PBH)}.\\ 
PBH are supposed to be formed in the very early universe during pre-stellar epoch, 
i.e.  prior to star formation .The idea of primordial black holes  and a possible mechanism
of their creation was pioneered by Zeldovich and Novikov~\cite{YaZ-IDN}.
According to their idea, the 
density contrast in the early universe inside the bubble radius, essentially equal to the cosmological horizon, 
might accidentally happen to be large, {${\delta \rho /\rho \approx 1}$,} then
that piece of volume would be inside its gravitational radius i.e. it became  a PBH, that
decoupled  from the cosmological expansion. \\
The mechanism was elaborated later by Hawking~\cite{SH-BH}, and by
Carr and Hawking~\cite{BC-SH}.

\end{enumerate}

\section{BH types by masses \label{s-BH-by-mass}}

Rather arbitrarly black holes are separated into three groups depending on their mass:\\
{1. Supermassive black holes (SMBH): $ M = (10^6 -10^{10}) M_\odot$.}\\
{2. Intermediate mass black holes (IMBH):  $ M = (10^2 -10^{5}) M_\odot$.}\\
{ 3. Solar mass black holes: masses from a fraction of $M_\odot$ up to $100 M_\odot$}.\\
4. There can be also very light black holes, not yet observed, with masses in the region $\sim10^{20}$ g;
they might be the "particles" of the cosmological dark matter. \\[1mm]
{The origin of most of these BHs is unclear, except maybe of the BHs with masses of a few
solar masses, which may be astrophysical}.\\
{Extremely unexpected was very high abundance of IMBH which are appearing during last several 
years in huge numbers.}\\
The assumption that (almost) all these black holes in the universe are primordial, 
except possibly the very light ones, strongly reduces or even
eliminates the tension between their observed abundances and possible mechanisms of their formation.
 
\section{Problems of the contemporary universe. Summary. \label{s-problems-sum}}

1. SMBH in  all large galaxies. The universe age is too short  for their formation through the commonly 
accepted accretion mechanism.  \\

2.  Several SMBH are found in very small galaxies and even in (almost) empty space, where not only
the time duration but also an amount of material is insufficient. 
An interesting recent observation was made by the Hobby-Eberly Telescope at Texas's McDonald Observatory 
suggesting the presence of a black hole with a mass of about 17 billion $M_\odot$ 
equivalent to 14\% of the total stellar mass of the galaxy.
Usually the mass of the central BH  is about 0.1 \% of the galaxy mass.
This SMBH was observed by the analysis of the motions of the stars near the center of the galaxy. 

There appeared recently fresh evidence~\cite{Leo-1} indicating to supermassive BH with
the mass $3\times 10^6 M_\odot$  in dwarf galaxy  Leo 1. Much more new data are presented practically 
today~\cite{dwarfs-SMBH}. Six dwarf galaxies are identified that  have X-ray AGN. They are presumbly
powered by SMBHs of  $M >  10^7 M_\odot $.

It is not excluded,  that such SMBHs, that are not hosted by a large galaxy, might be
pushed out of large galaxies in the process of galaxy collisions. Such catastrophic event may even create 
plenty of wandering single supermassive black holes. However, taking into account a large number of
such exotics, much more natural seems that all SMBH in small galaxies are primordial. Simply they were
unlucky not to acquire their own large galaxy, since there was not enough matter around to build large
galaxies.

3. Prediction of abundant population of BHs with masses $\sim 10 M_\odot$ in the Galaxy,
not yet observed.  

4. Origin and properties of the sources of the observed gravitational waves, encounter 
considerable difficulties, if 
one tries to explain them assuming astrophysical formation of back hole binaries emitting the 
gravitational radiation.

5. IMBH, with ${M \sim (10^3-10^5) M_\odot}$ are unexpectedly discovered in dwarfs and globular clusters.
Their origin is unclear, if they are not primordial.

6. Invisible Massive Astrophysical Compact Halo Objects (MACHOs), non-luminous objects with masses  
${\sim 0.5 M_\odot}$ observed  through microlensing. It is unknown what are they and how they were 
created.

7. Existence of very unusual stars in the Galaxy, among which there are too fast moving stars and  
stars with unusual chemistry. Moreover, too old stars, are found. Many of them look older than the 
Galaxy and maybe one is even older that the universe (sic!?).

An assumption, that the black holes mentioned in the list above, are primordial eliminates all the problems.
The mechanism of PBH formation suggested in papers~\cite{AD-JS,DKK} predicts also existence
of  the unusual stars mentioned in point 7.


\section{Observations of black holes \label{s-BH-obs}}

{The ancient point of view is that BH are objects with so strong gravitational field that nothing can escape it.}
According to Mitchell (1784): there may be bodies for which the second cosmic velocity is larger 
than the speed of light.
They do not shine and do not reflect light, i.e. are absolutely dark,  invisible.

However, the truth is quite the opposite, black holes are very well seen. 
Light BHs can emit all kind of radiation through  the Hawking evaporation 
(though nobody has yet seen it).
{The most powerful sources of radiation in the universe 
are SMBH - quasars, point-like objects shining as a thousands of
galaxies.}\\[1mm]
The methods of the BH observations include:\\
1. Central mass estimate through analysis of stellar motion around the supposed BH
as e.g. discovery of BH in the center of the Millki Way. \\
2. Distortion of star motion due to invisible point-like gravitating body.\\ 
3. Gravitational lensing (MACHO and some other BHs).\\
 4. Electromagnetic radiation from the accreting matter; it is the mechanism of quasar central engine, but
 much smaller BHs are also observed that way.\\[1mm]
However,  all these methods allow only to establish that there is a large mass inside small volume.
We need theory to proceed further and to conclude  that there should  be  a black hole  inside.  
But the following method is free  from this restriction:\\ 
4. Registration of gravitational waves from coalescing double systems of black holes. The data directly
show that there are exactly coalescence of two BHs. This is the first test of General Relativity for strong fields 
and the first observational proof of existence of the Schwarzschild solution.

\section{PBH and inflation \label{s-prop}}

The mechanism suggested in ref.~\cite{AD-JS,DKK} introduced some new 
features which were later explored in a series of subsequent works. 
The proposed there scenarios
are heavily based on the Affleck-Dine~\cite{AD} model
of baryogenesis, that 
permits to create very interesting features of the PBH population or some other macroscopic compact
objects, see below, sec.~\ref{s-pbh-AD}   

In paper~\cite{AD-JS}  inflationary
mechanism was first implied for PBH formation. It allowed
to  create PBH with huge masses, much larger than those in the previously studied models.
A year later inflationary creation of PBH was  explored in ref.~\cite{infl-2},  soon after that in ref.\cite{infl-3},
and two years later in~\cite{infl-4}.
Nowadays there is  an avalanche of 
papers on inflationary formation of PBH.

However, except for predicting large masses of PBH, the models do not have much predictive
power because the mass spectra of the created PBHs are quite complicated and strongly
parameter dependent. No simple analytic expressions have been presented. The only exception
is the mechanism of refs~\cite{AD-JS,DKK}, which predicts extremely simple log-normal
mass spectrum of PBH:
\be
\frac{dN}{dM} = \mu^2 \exp{[-\gamma \ln^2 (M/M_0)] }.
\label{dn-dM}
\ee
The central value mass can be calculated theoretically~\cite{AD-KP}:
${M_0 \sim 10 M_\odot}$. It is equal to the horizon mass at QCD phase transition  from the phase
of free quark-gluon plasma to the confinement phase. To be more precise the  horizon mass  is
approximately equal to $10M_\odot$ for the cosmic plasma with vanishingly small chemical potential 
$\mu$. In our case 
$\mu $ is supposed to be large,\ of the order of the plasma temperature. Correspondingly the phase transition
was probably delayed and the horizon mass could be somewhat bigger. 

An impressive feature of the the {log-normal} mass spectrum with the predicted value  of $M_0$ is that
it is the only known  spectrum tested by "experiment"\, in very good agreement  with the observed 
densities of black holes in all mass intervals from the solar mass BH, up to black holes with intermediates
masses, and further up to supermassive black holes. In particular,
the mechanism developed in~\cite{AD-JS,DKK} allows to explain the
presence of SMBH in all large and several small galaxies accessible to observation.
For very massive BH an account should be taken 
of the mass rise due to later matter accretion.

Especially impressive is the confirmation of the model by the 
chirp mass binaries measured by LIGO/Virgo which is discussed in section~\ref{GW-BH}.

\section{Black Dark Matter \label{black-DM}}

The first suggestion PBH might be dark matter "particles"  was made by
S. Hawking in 1971~\cite{SH-BH}
and later by G. Chapline in 1975~\cite{chaplin} 
who noticed that low mass PBHs might be abundant 
in the present-day universe with the density comparable to the density of dark matter. 
The scale independent spectrum of cosmological perturbations was assumed. That
led to the flat mass spectrum in log interval:
\be
 dN = N_0 (dM/M) 
\label{flat=spectrum}
\ee
with maximum mass { ${ M_{max} \lesssim 10^{22}  }$ g,} which hits the allowed mass range.

The next paper on PBH made dark matter belongs  to A. Dolgov, J. Silk  (Mar 13, 1992)~\cite{AD-JS}
that predicted much larger PBH masses. It was the first paper where inflation was
applied to  PBH formation, so PBH masses as high as 
$ 10^6 M_\odot$, and even higher, can be created. 
The simple log-normal mass spectrum of PBH was predicted.

The constraints on the cosmological mass density of black holes are reviewed in two papers~\cite{CK1,CK2}
for monochromatic and extended (in particular log-normal)
mass spectrum of PBHs.  As it was mentioned B. Carr in 2019
all limits are model dependent and have caveats. The summary plot on PBH density limit is presented 
in Fig. 1.
\begin{figure}[htbp]
\begin{center}
\label{CK-limits}
\includegraphics[scale=0.5]{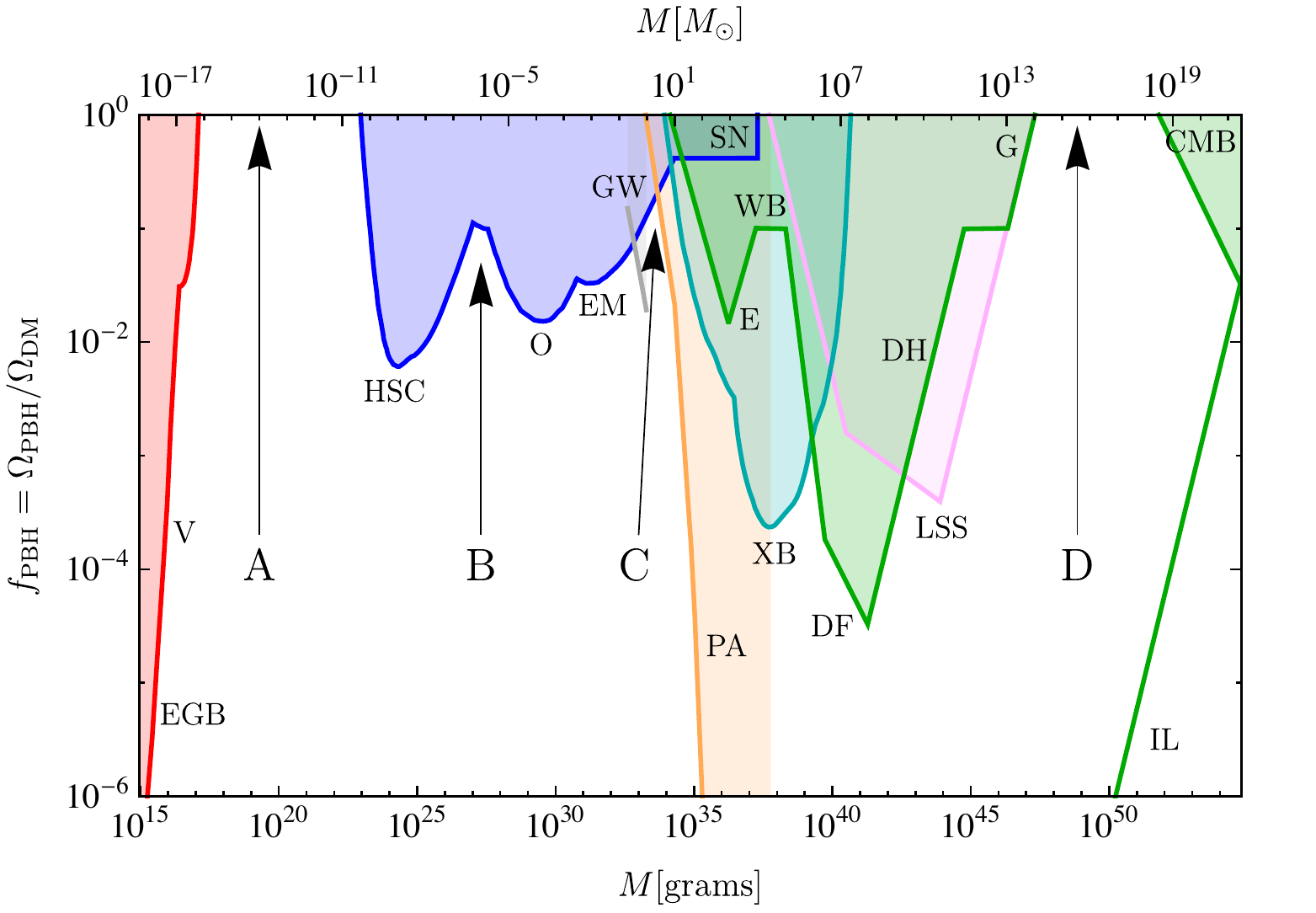}
\caption{Constraints on $f( M )$ 
			for a  monochromatic
			mass function,
			There are four mass windows (A, B, C, D) 
			in which PBHs could have an appreciable density. }
			
\end{center}
\end{figure}


The arguments presented in ref.~\cite{boehm} permit to weaken the constraints 
on the PBH density in the mass range
$(30-100) M_\odot$ and to reopen the door for dark matter in the form of PBHs registered by LIGO.
The point is that
PBHs were treated as point Schwarzschild masses, while the more careful analysis in an expanding universe presented in this work, leads to a time-dependent mass. This implies a stricter set of conditions for a black hole binary to form and means that black holes coalesce much more quickly than was previously calculated, namely well before the LIGO/Virgo's observed mergers. The observed binaries are those coalescing within galactic halos, with a merger rate consistent with data. This opens the door
 for dark matter in the form of LIGO-mass PBHs.

The bounds presented in \cite{CK1,CK2} for the intermediate mass black holes were criticised 
also in ref.~\cite{cor-fram}. 
The most questionable step in this chain of arguments is the use of overly simplified accretion models. 
The same accretion models was applied to X-ray observations from supermassive black holes, M87 and 
Sgr $A^*$. The comparison of these two SMBHs with 
 intermediate mass MACHOs suggests that the latter could, after all, provide a significant constituent of all 
the dark matter.

One more argument in favour of allowed large cosmological density of PBH is based on the possibility that
PBHs can form clusters, as it is argued in ref.~\cite{SRK}. 
Dynamical interactions in PBH clusters offer additional channel for the orbital energy dissipation increase 
the merging rate of PBH binaries, and the 
{constraints on the cosmological fraction of BH in DM, $f_{PBH}$, made by these black holes,
that were obtained by assuming a homogeneous PBH space distribution can be weaker.} 
 A recent analysis performed in~\cite{Er-St} based on the PBH formation 
 model ~\cite{sasaki} and \cite{nakamura} shows that even $f_{PBH} = 0.1 - 1$ is not 
 excluded \footnote{I thank K. Postnov for indicating these references.}.

So the presented in the literature strong bounds on the cosmological density of PBH should be taken with a 
grain of salt.

\section{Gravitational waves from BH binaries \label{GW-BH}} 

There is general agreement between several groups that the gravitational waves discovered by
LIGO/Virgo interferometers originated from PBH binaries. We discuss this issue  here 
following our paper~\cite{BDPP}. There are three problems which indicate that the sources of
GWs are most naturally primordial black holes:\\
1. Origin of heavy BHs  (with masses ${\sim 30 M_\odot}$). To form so heavy BHs, the progenitors should have 
${M > 100 M_\odot}$ and  a low metal abundance to avoid too much
mass loss during the evolution. Such heavy stars might be present in
young star-forming galaxies but they are not observed in the necessary amount.
Recently there emerged much more striking problem because of the observation
of BH with ${M \sim 100M_\odot}$. Formation of such black holes 
in the process of stellar collapse was considered  to be strictly  forbidden.
Some exotic mechanisms might be possibly allowed, such as e.g. BH formation
in the process of collapse of supermassive star heated by dark matter 
annihilation inside~\cite{ZF}. \\
On the  other hand, primordial black holes with the observed by LIGO masses may be easily created 
with sufficient density. \\
{2. Formation of BH binaries from the original stellar binaries. }
Stellar binaries are
formed from common interstellar gas clouds and are quite frequent in galaxies.
If BH is created through stellar collapse,  small non-sphericity results in a huge 
velocity of the BH and the binary is destroyed.
{BH formation from PopIII stars and subsequent formation of BH
binaries with  tens of ${M_\odot}$
is estimated to be small.}
{The problem of the binary formation is simply solved if the observed sources of GWs are the binaries of
primordial black holes.} 
{They were at rest in the comoving volume, when inside horizon they were gravitationally attracted and  
might loose energy due to dynamical friction in the early universe.
The probability for them to become gravitationally bound is significant.}\\
The conventional astrophysical 
scenario is not excluded but less natural.\\
{3.  Low spins of the coalescing BHs .}
{The low values of the BH spins sae observed
in GW150914 and in almost all (except for three) other events.}
It strongly constrains astrophysical BH formation from close binary systems. 
{Astrophysical BHs are expected to have considerable angular momentum, nevertheless the
dynamical formation of double massive low-spin BHs in dense stellar clusters is not excluded, though difficult.} 
{On the other hand, PBH practically do  not rotate because vorticity perturbations 
in the early universe are vanishingly small.}
{Still, individual PBH forming a binary initially rotating on elliptic orbit could gain collinear spins 
about 0.1 - 0.3, rising with the PBH masses and eccentricity~\cite{Post-Mit,PKM}.
This result is in agreement with the
GW170729 LIGO event produced by the binary with masses ${50 M_\odot}$  and ${30 M_\odot}$ and
and  GW190521.

To summarise: each of the mentioned problems may be solved in the conventional frameworks but it looks
much simpler to assume that the LIGO/Virgo sources are primordial.

\section{Chirp mass distribution \label{s-chirp}}

Two rotating gravitationally bound massive bodies are known to emit gravitational 
waves, as is discussed in the previous section.
In quasi-stationary inspiral regime, the radius of the orbit and the rotation frequency
are approximately constant and the GW frequency is twice the rotation frequency. 
{The luminosity of the GW radiation is:}
\be 
L  = \frac{32}{5}\,m_{Pl}^2\left(\frac{M_c\,\omega_{orb}}{m_{Pl}^2}\right)^{10/3}\,,
\label{L-gW}
\ee
where $M_1$, $M_2$ are the masses of two bodies in the binary system and 
${M_c}$ is the so called chirp mass: 
\be 
M_c=\frac{(M_1\,M_2)^{3/5}}{(M_1+M_2)^{1/5}} \, ,
\label{M-c}
\ee
and 
\be 
\omega^2_{orb} =  \frac{M_1+M_2}{m_{Pl}^2 R^3}\,.
\label{omega}
\ee
 
In  ref.~\cite{chirp} the available data on the chirp mass distribution of the black holes in the 
coalescing binaries in O1-O3 LIGO/Virgo runs are
analyzed and compared with theoretical expectations based on the hypothesis that these black 
holes are primordial with log-normal mass spectrum.
{The inferred best-fit mass spectrum parameters are: }
{$M_0=17 M_\odot$ and ${\gamma=0.9}$,} fall 
within the theoretically expected range and show excellent agreement with observations. 
{On the opposite, binary black hole formation based }
{ on massive binary star evolution} require additional adjustments to 
reproduce the observed chirp mass distribution. 
The results are presented in Figs. 2 and 3.

\begin{figure}[htbp]
\begin{center}
\includegraphics[scale=0.2]{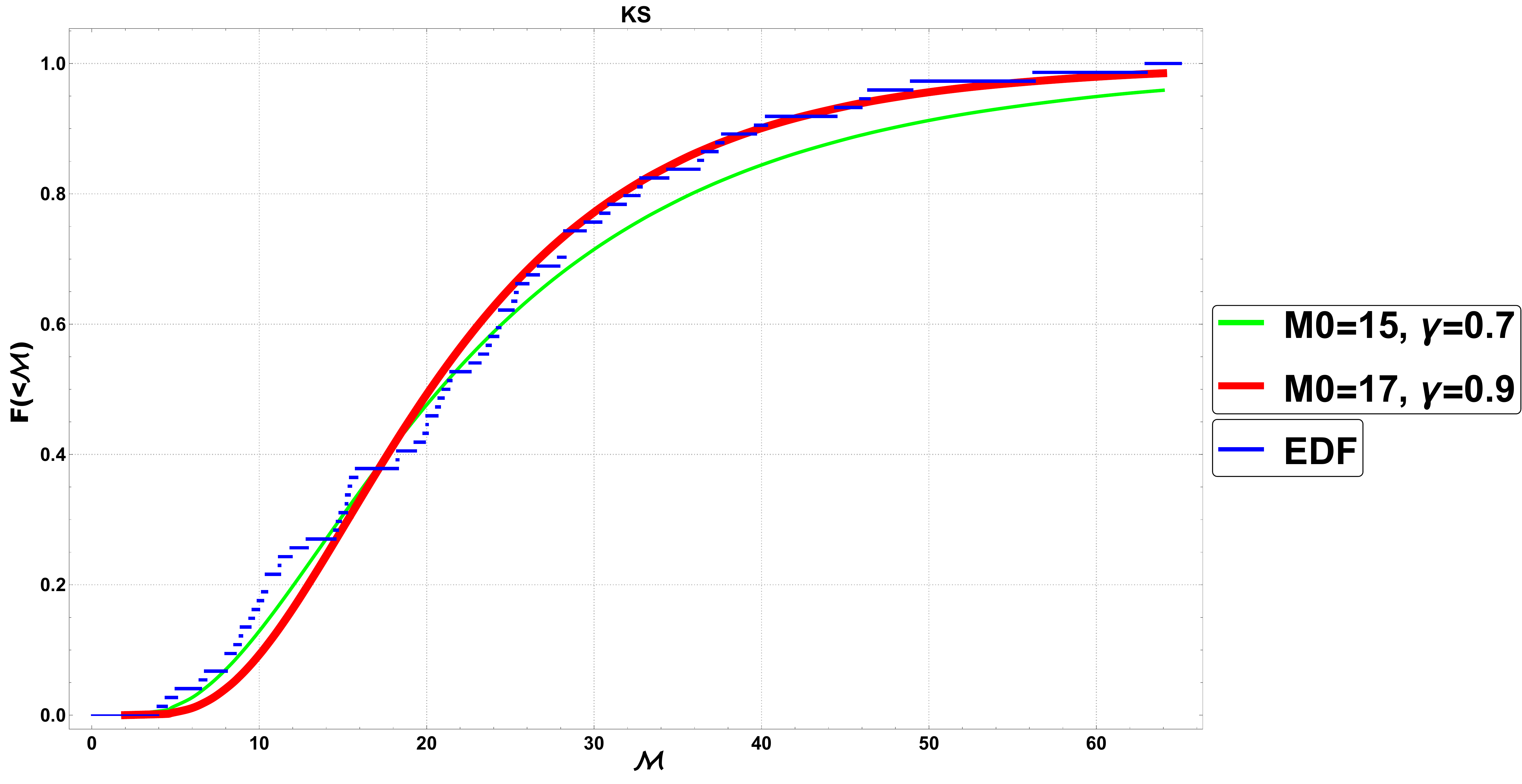}
\caption{Model distribution $F_{PBH}(< M)$ with parameters  $M_0$ and ${\gamma}$ for two best 
Kolmogorov-Smirnov tests.  EDF= empirical distribution function.}
\end{center}
\end{figure}

\begin{figure}[htbp]
\begin{center}
\includegraphics[scale=0.20]{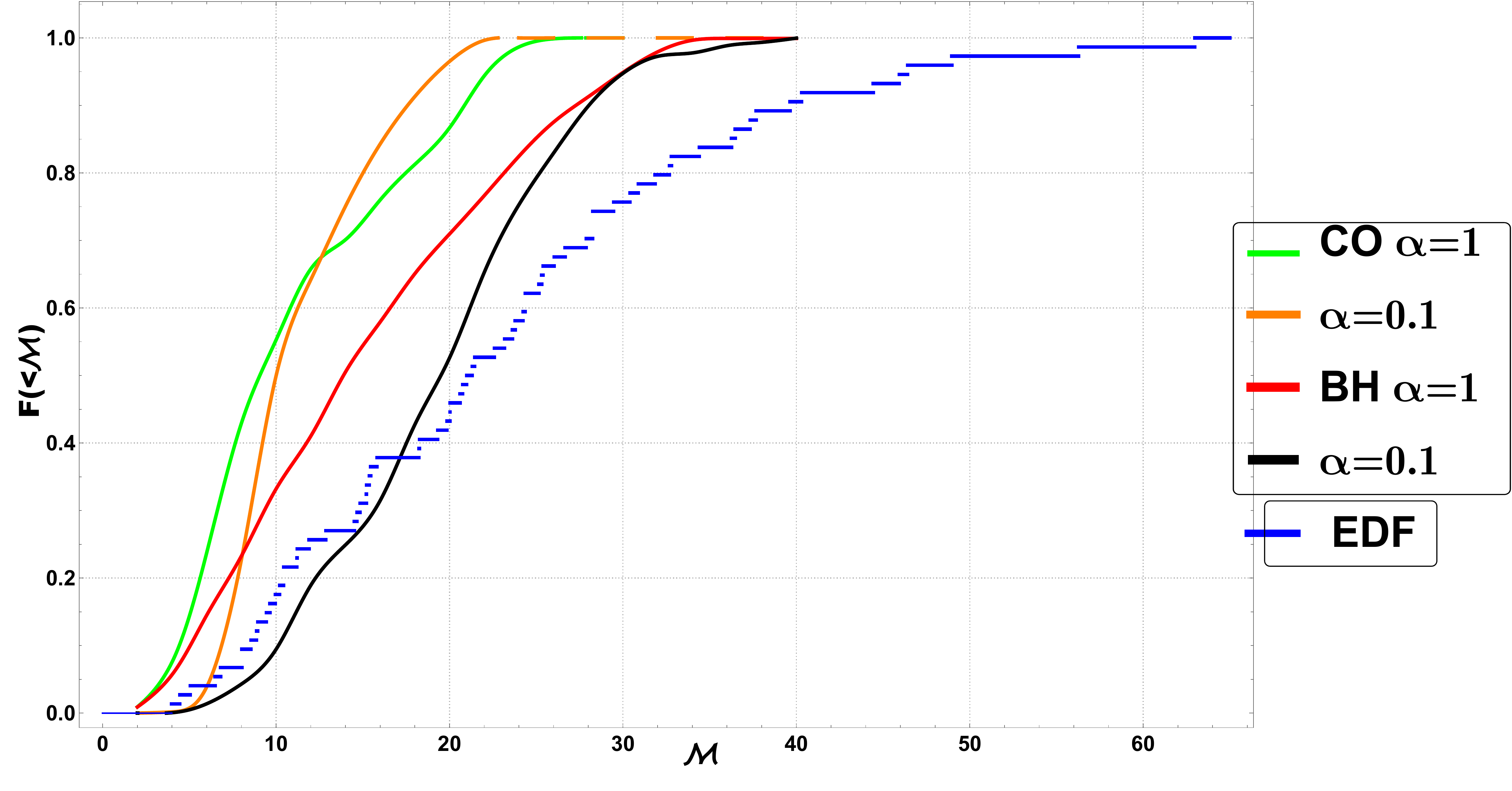}
\caption{Cumulative distributions $F(< M)$ for several { astrophysical} models of binary BH coalescences. }
\end{center}
\end{figure}

So we can conclude that PBHs with log-normal mass spectrum perfectly fit the data, while 
astrophysical BHs seem to be disfavoured.

New data on GW observations were analysed by K.A. Postnov in his talk at
XXXIV International Workshop on High Energy Physics "From Quarks to Galaxies: Elucidating Dark Sides"
are depicted in fig. 4. 

In this talk it is also presented an approximate fitting 
of the observed chirp-mass distribution in the O1-O3 LVK GW compact binary
coalescences (from GWTC-3 catalog) by two independent PBH populations with initial log-normal mass 
distributions $M_0^{(1)} = 5 M_\odot $ and $M_0^{(2)} = 30 M_\odot $, see fig. 5

\begin{figure}[htbp]
\begin{center}
\includegraphics[scale=0.35,angle=-90]{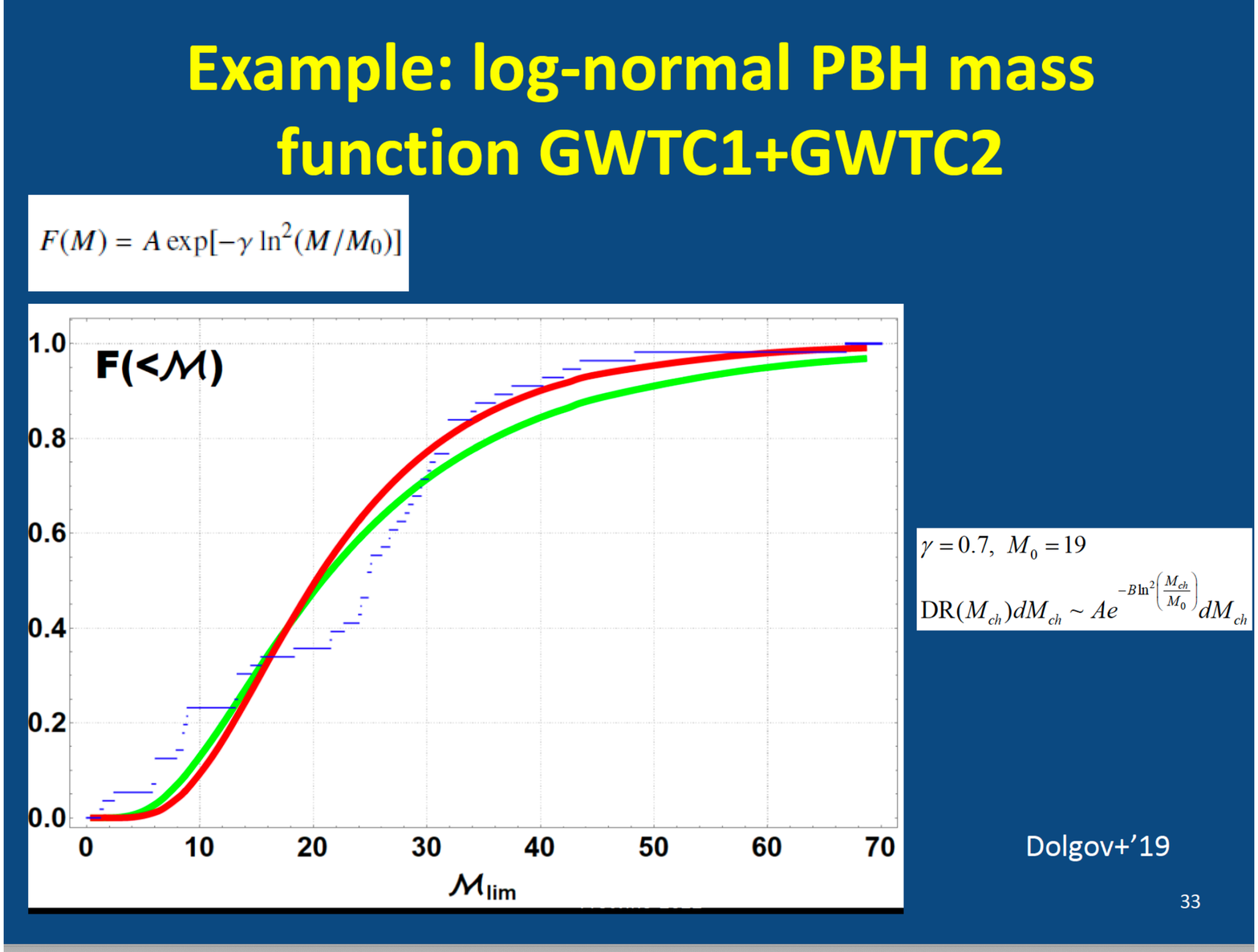}
\caption{}
\end{center}
\end{figure}

\begin{figure}[htbp]
\begin{center}
\includegraphics[scale=0.20]{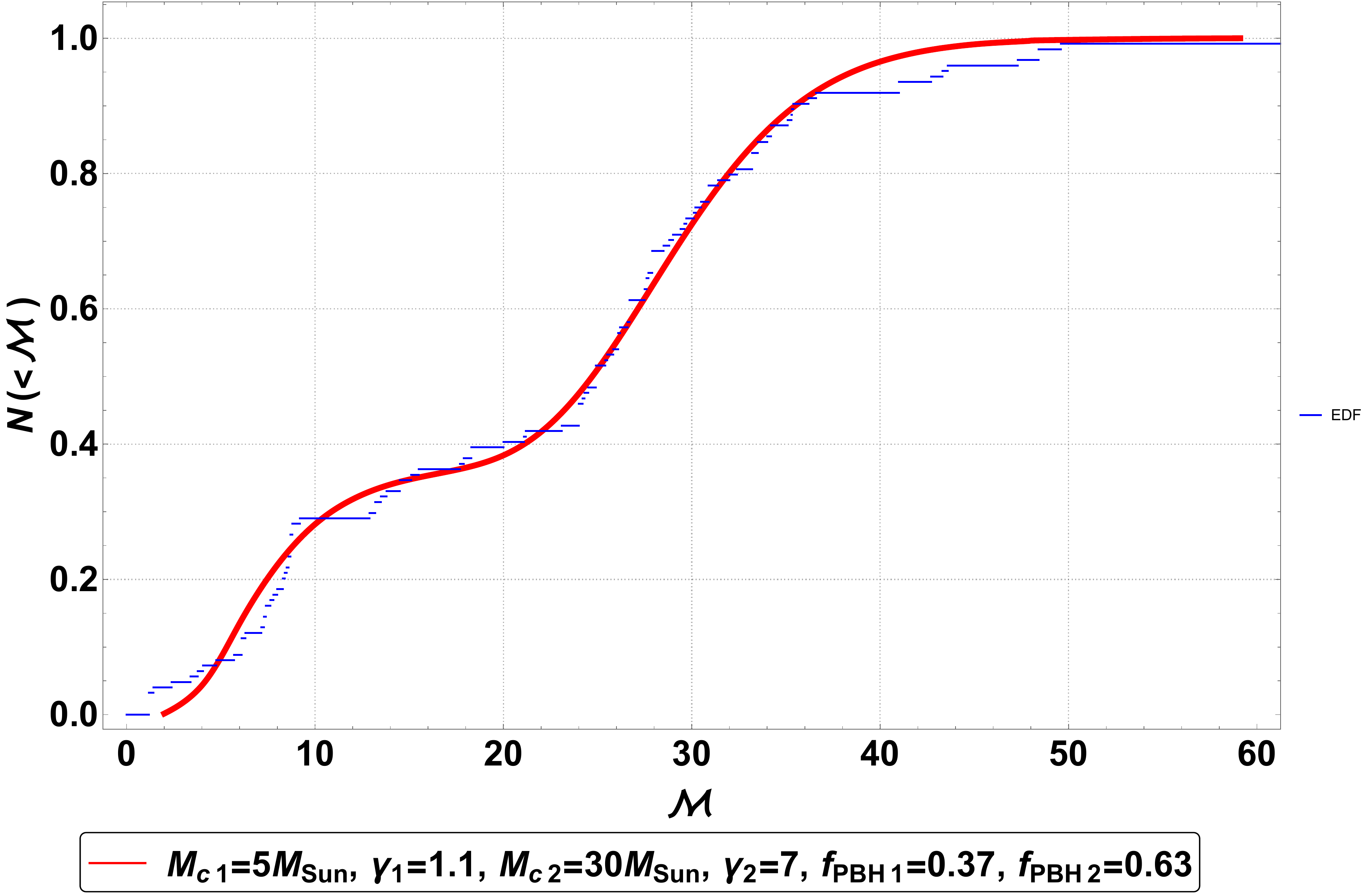}
\caption{Approximation of the observed chirp-mass distribution in the O1-O3 LVK GW compact
binary coalescences (from GWTC-3 catalog) by two independent PBH populations, from K.Postnov talk}
\end{center}
\end{figure}

In fig. 6 one can see the same approximation but
 by the simplest model of astrophysical BH formation from the collapse of
the CO-core of a massive star and standard common-envelope parameter,
with taking into account evolution of star-formation rate in the universe with redshift 
plus a population of PBHs with log-normal initial mass 
distribution with $M_0^{(2)} = 33 M_\odot $. 

\begin{figure}[htbp]
\begin{center}
\includegraphics[scale=0.35]{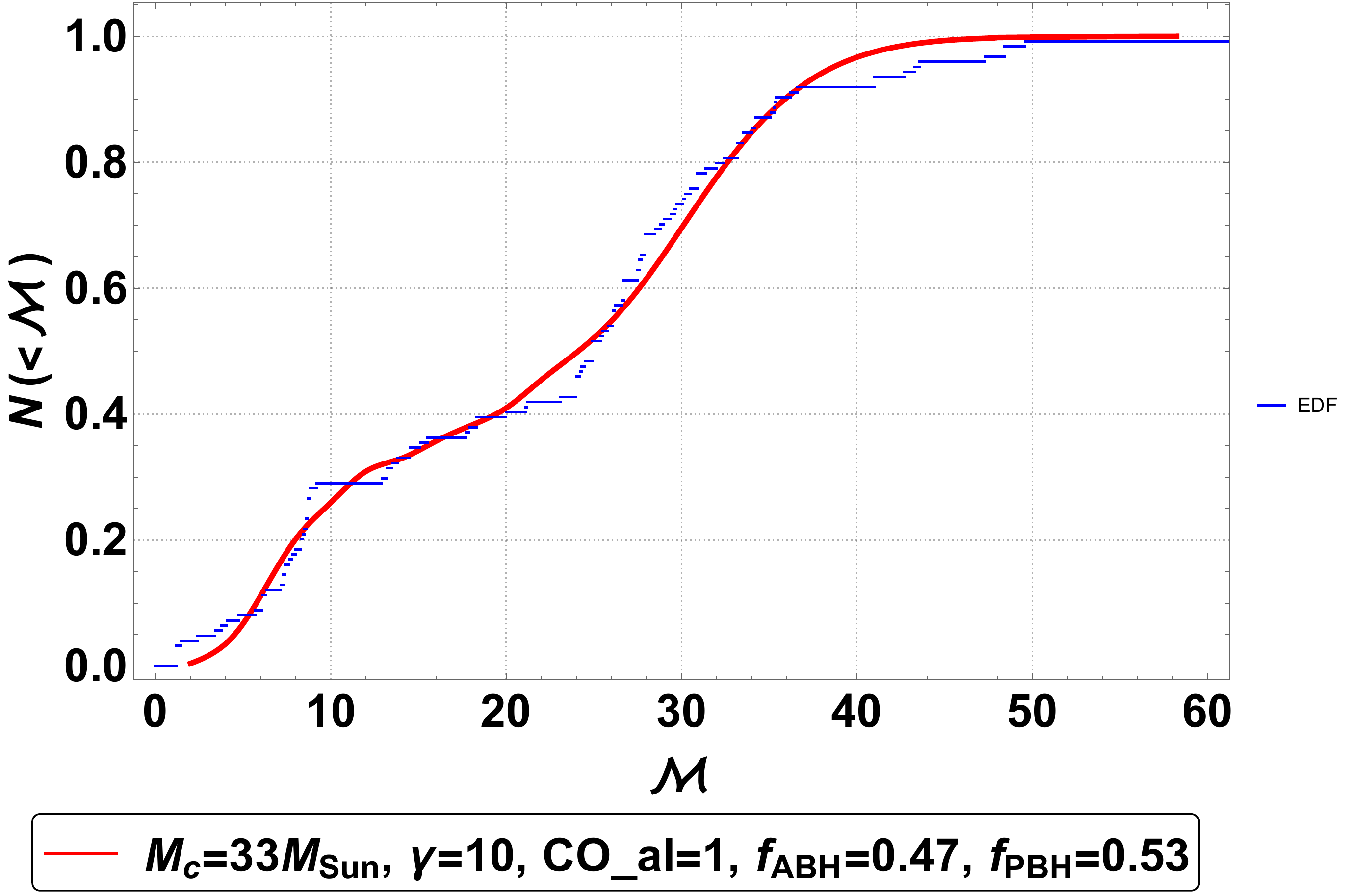}
\caption{The simplest model of astrophysical BH formation from the collapse of
the CO-core of a massive star, from K.Postniv talk}
\end{center}
\end{figure}

The picture looks more complicated than the earlier one described at the beginning of this section.
One possible interpretation is that there are two populations of PBH with log-normal mass spectrum each
but with different values of $M_0$. The model can be modified to allow that and it was even discussed earlier
in the author's papers, but aesthetically it looks not so nice. Another option is that there a mixture of primordial
and astrophysical black holes observed by LIGO/Virgo. and even a possibility that some binaries consist of 
a pair of primordial and astrophysical black holes. Hopefully future data will help to resolve all the
controverses. 

\section{Dwarfs and globular clusters \label{s-dwarfs}}

A large number of intermediate mass black holes,
that were discovered during last decade, hardly fits the narrow 
frameworks of the standard cosmological model. However, if they are primordial with the 
determined above parameters 
of the log-normal mass spectrum, their number is just what is necessary to explain the data and in particular
to understand the mechanism of the origin of dwarf galaxies and  globular clusters which is not well understood 
in the conventional cosmology. Recently possible discovery of SMBH in dwarf galaxy Leo 1 was 
announced~\cite{Leo-1}. Such SMBH surely could not be created by the galactic matter
accretion to  the galaxy centre. 

Much more new data are presented practically 
today~\cite{dwarfs-SMBH}. Six dwarf galaxies are identified that  have X-ray AGN,
powered by SMBHs of  $M >  10^7 M_\odot $. Most probably these dwarfs were seeded by
primordial supermassive black holes in accordance with ref~\cite{AD-KP-glob}.

As argued in ref.~\cite{AD-KP-glob}
IMBHs with masses of a few thousand solar mass, or higher, can seed
 formation of globular clusters (GCs) and dwarf galaxies.
 In the last several years such IMBH inside GSs are observed confirming this suggestion. 
 For example  the BH  with the mass $M \sim 10^5 M_\odot$ is discovered almost yesterday
 in the dwarf galaxy SDSS J1521+1404. 

The astrophysical origin of IMBH encounter serious problems
for all mass values, but nevertheless huge number of them 
are discovered with all possible masses.
On the contrary, the described above model of PBH formation excellently resolves all the inconsistencies.

\section{Intermediate summary and antimatter in the Galaxy}

The mechanism of PBH formation suggested in refs~\cite{AD-JS, DKK} neatly cure
all the problem related to the observed population of the universe at high redshifts as
well as of the present day universe;
{The predicted log-normal spectrum of PBH is tested and confirmed by the observations 
(the only one existing in the literature).
{The predicted existence of IMBH in globular
clusters is confirmed.} \\[1mm]
 So the model works great.}
 Thus the seemingly crazy by-product of refs~\cite{AD-JS,DKK}, namely prediction of antimatter in the Galaxy
 can come true as well. Probably it is indeed the case.
 A surprisingly huge flux of cosmic positrons, of He-antinuclei, and possibly even a 
population of antistars seem to be observed.

\subsection{Anti-evidence: cosmic positrons \label{ss-positrons}}

The observation of intense 0.511 line see refs~\cite{pos-1,pos-2,pos-3}, and earlier references therein,
presents  a strong proof of abundant positron population in the Galaxy.
 In the central region of the Galaxy 
electron--positron annihilation proceeds  at a surprisingly high rate, creating the flux: 
\be 
\Phi_{511 \; {\rm keV}} = { 1.07 \pm 0.03 \cdot 10^{-3} }\; 
{\rm photons \; cm^{-2} \, s^{-1}} .
\label{pos-flux}
\ee
The width of the line is about 3 keV. It proves that the annihilation takes place at rest.
The emission mostly goes from the  Galactic bulge and  at much lower level from the disk,
The source of 0.511 MeV line   in the Galactic bulge. even got the name "Great Annihihilator".

Until recently the commonly accepted explanation was that  ${e^+}$ 
are created in the strong magnetic fields of pulsars but the recent results of AMS probably exclude 
this mechanism, since the spectra of $\bar p$ and $e^+$ at high energies are identical~\cite{Ting-1,Ting-2}. 
It means that their
origin is the same and since antiprotons are not created  in magnetic fields of pulsars, the conclusion
might follow that positrons are not produced by pulsars
as well. However, this might be true only for positrons with 
very high energies, about or above tera-electronvolts.

\subsection{Anti-evidence: cosmic antinuclei \label{ss-anti-nuc}}

The striking result reported by AMS team is the registration of anti-Helium-3 and anti-Helium-4 nuclei. Namely
in 2018 AMS-02 announced possible observation of six
${\overline{He}^3}$ and two ${\overline{He}^4}$~\cite{Ting-1}.
In 2022 already 7 $\overline D$ ($E \lesssim 15$ GeV) and 9 $\overline{He}$, 
($E \sim 50$ GeV) were  observed. Surprisingly high 
fraction $\overline{He}/He \sim10^{-9}$ was registered~\cite{Ting-2,anti-He-1}.
It is not excluded that the flux of anti-helium is even much higher because low energy 
${\overline{He}}$ may escape registration in AMS.



Secondary production of different antinuclei in cosmic ray was estimated  in ref.~\cite{cosm-anti-nuc}.
According to this work  anti-deuterium could be  
most efficiently
produced in the collisions ${\bar p\,p}$ or
${\bar p\, He}$ which can create the flux
${\sim 10^{-7} /m^{2}/ s^{-1}} $/steradian/GeV/neutron),
i.e. 5 orders of magnitude below the observed flux of antiprotons.
Antihelium could be created in the similar reactions and 
the fluxes of  ${\overline{He}^3}$ and ${\overline{He}^4}$, that
could be created in 
cosmic rays would respectively be 4 and 8
orders of magnitude smaller than the flux of anti-D.


After the AMS announcements of observations of  anti-$He^4$  there
appeared theoretical attempts to create 
anti-$He^4$ through dark matter annihilation. This possibility does not look natural. Moreover,
DM annihilation would presumably create strong cosmic ray flux of other particles, which is not observed.

 In accordance with our model the observed antinuclej can be signatures of primordial antimatter. However 
 if they are synthesised in the standard big bang anti-nucleosynthesis (anti-BBN) one would naturally expect the same
 abundances  of light elements as those created by the canonical BBN. According to the latter the abundances of 
 deuterium and helium-3 are much smaller then that of helium-4, approximately by 4 orders of magnitude,
 while the relative fraction of these antinuclej are approximately equal. There might be some astrophysical
 explanation of that or this anomaly is related to the fact that in our model antismatter is created in bubbles
 with unusually high baryon-to-photon ratio $\beta$. In the canonical BBN $\beta \sim 10^{-9}$, while in our
 case it may be as large as unity. 
However if $\beta \sim 1$ there is no primordial D. On the other hand in our scenario  the formation of primordial 
elements takes place inside non-expanding compact stellar-like objects with fixed temperature. If the temperature is sufficiently high, this so called BBN may stop before abundant He formation  and ends 
with almost equal abundances of D and He. One can see that looking at 
abundances of light elements at a function of temperature. 
If it is so, antistars may have equal amount of 
$\overline{D}$ and $\overline{He}$.

\subsection{Anti-evidence: antistars in the Galaxy \label{ss-anti-star}}

Almost two years ago a sensational announcement was done~\cite{anti-stars}
 about possible discovery of 14 antistars in our Galaxy, Milky Way.
Quoting the authios:  "We identify in the catalog 14 antistar candidates not associated with any objects belonging 
to established gamma-ray source classes and with a spectrum compatible with 
baryon-antibaryon annihilation''. The map of the observed anti-sources is presented in fig.~7.

\begin{figure}
\includegraphics[scale=0.6]{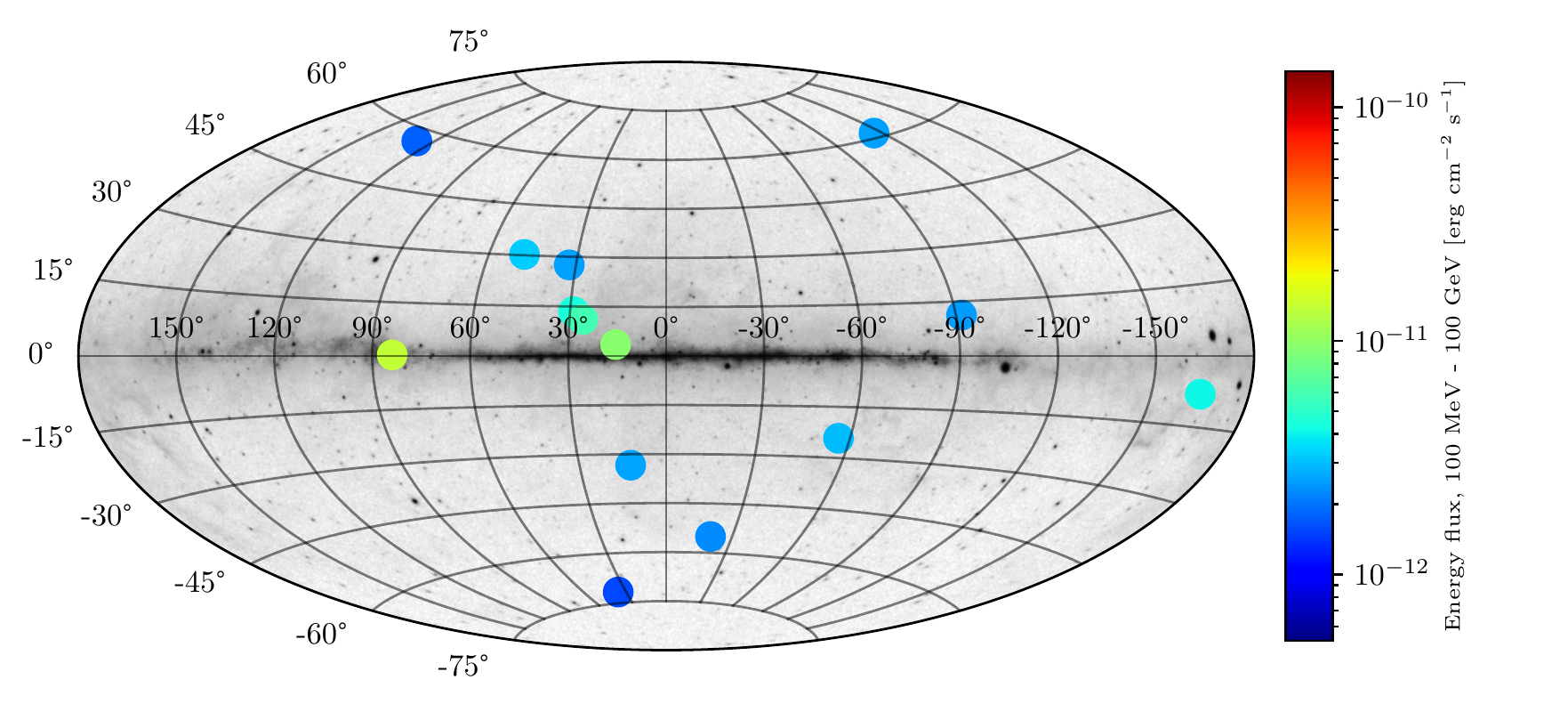}
\caption{\label{fig:sources} Positions and energy flux in the 100 MeV - 100 GeV range of antistar candidates 
selected in 4FGL-DR2. Galactic coordinates. The background image shows the Fermi 5-year all-sky photon 
counts above 1 GeV 
}
\end{figure}  

An additional possible method for antistar detection in the Galaxy or in its halo has been proposed in
ref.~\cite{BBBDP}.
In astrophysically plausible cases of the interaction of neutral atmospheres or winds from 
antistars with ionised interstellar gas, the hadronic annihilation 
 will be preceded by the formation of excited $ {p \bar p}$
and $He {\bar p}$ atoms. These atoms rapidly cascade down to low levels prior to 
annihilation giving rise to a series of narrow lines which can be associated with the hadronic 
annihilation gamma-ray emission. The most significant are L (3p-2p) 1.73 keV line (yield more 
than 90\%) from ${p \bar p}$ atoms, and M (4-3) 4.86 keV (yield $\sim 60$\%) and L (3-2) 11.13 
keV (yield about 25\%) lines from $He^4 \bar p$ atoms. These lines can be probed in dedicated 
observations by forthcoming sensitive X-ray spectroscopic missions XRISM and Athena and in 
wide-field X-ray surveys like SRG/eROSITA all-sky survey.

Bounds on the possible density if antistars in the Galaxy were studied in several 
papers~\cite{lim-1, lim-2, lim-3}. It was shown that the restrictions are rather mild and
 an abundant density of compact
anti-stars in the universe even in the Galaxy does not violate existing observations.  
The reason is that the annihilation proceeds on a thin surface layer with a very short depth 
of the order of the proton mean free path in the dense stellar medium. On the other hand
if there were disperse  antimatter clouds, the annihilation would be by far more efficient and if so, the
anticlouds did not survive to our time, though they might have been existed in the early universe 

A vey impressive would be star-antistar collision, which may even be a quasi-periodic process of a 
star-antistar direct contact, explosion forcing them apart and return to each other by gravitational
 attraction, etc...

\section{PBH and anti-creation mechanism \label{s-pbh-AD} }

The model of PBH and antimatter creation of refs.~\cite{AD-JS,DKK}
 is essentially based on the supersymmetry (SUSY) motivated baryogenesis, 
 proposed by Affleck and Dine (AD)~\cite{AD},
 though the full extend SUSY is not necessary.
 SUSY predicts existence of  scalar field $\chi$ (or several such fields)  with non-zero baryon number, $ B\neq 0$.
 Another important feature of the scenario is existence of the  flat directions in the self-potential of $\chi$, i.e. the
 directions along which the potential does not rise. Simple examples of such quadratic and quartic potentials with
 flat directions are the following, for quartic self-interaction
\be
U_\lambda(\chi) = \lambda |\chi|^4 \left( 1- \cos 4\theta \right)
\ee
and for the quadratic mass-like term: 
\be
U_m( \chi ) = m^2 |\chi|^2 [ [1-\cos (2\theta+2\alpha)] 
\ee
where ${ \chi = |\chi| \exp (i\theta)}$ and ${ m=|m|e^\alpha}$.
{If ${\alpha \neq 0}$, C and CP are  broken.} 
In GUT SUSY baryonic number is naturally non-conserved.
In our toy model it is described by  non-invariance of ${U(\chi)}$
with respect to phase rotation, $\chi \rar \chi \exp (i \theta) \chi $. 

In the course of inflation $\chi$ quantum-fluctuates along flat directions with increasing amplitude 
due to quantum instability of massless fields at De Sitter stage~\cite{ADL-ins,AAS-ins}. 
Thus taking into account that the wave length of the fluciuations exponentially rises, 
$\chi$ could  effectively acquire a large classical value.

When  inflation is over and the symmetry maintaining flat directions brakes down, $\chi$
starts to evolve to the equilibrium point, ${\chi =0}$,
according to the equation of the Newtonian mechanics with the liquid friction (Hubble friction) term:
\be
\ddot \chi +3H\dot \chi +U' (\chi) = 0.
\ee
Due to quantum fluctuations orthogonal to the flat directions, $\chi$ obtains momentum in 
the orthogonal to the valley direction.
This is how baryonic number of $ \chi$ is generated:
\be
B_\chi =\dot\theta |\chi|^2,
\ee
$B$ is analogous to mechanical angular momentum in the two dimensional complex plane 
$ [{\Re}e \chi, {\Im}m\, \chi] $.

Decays of $\chi$ into quarks and antiquarks is supposed to conserve baryonic number and transforms
the baryonic  number of $\chi$ into the baryonic number of quarks. In this process a huge cosmological
baryon asymmetry can be generated, much larger than the observed one, 
$\beta \approx 10^{-9}$. 
If $m \neq 0$, the angular momentum, B, could be generated due to possible different 
direction of the  quartic and quadratic valleys at low ${\chi}$. In this case orthogonal quantum 
fluctuation are unnecessary.
{If CP-odd phase ${\alpha}$ is small but non-vanishing, both baryonic and 
antibaryonic domains might be  formed}
{with possible dominance of one of them.}
 
In the model of ref.~\cite{AD-JS, DKK}  the AD, scenario of baryogenesis was essentially modified 
by an addition
of interaction of the Affleck-Dine field $ \chi$ with the inflaton,  $\Phi$. The interaction potential is taken in
the form:
\be 
U = g|\chi|^2 (\Phi -\Phi_1)^2 +
\lambda |\chi|^4 \,\ln \left( \frac{|\chi|^2 }{\sigma^2 } \right)
+(\lambda_1 \chi^4 + h.c. ) + 
(m^2 \chi^2 + h.c.). 
\label{U-tot}
\ee
The first term in this expression is the new type of interaction potential between $\Phi$ and $\chi$.
$\Phi_1 $ is a constant. It is the 
amplitude of the inflaton field,  taken by $\Phi$ in the process of inflation. The remaining duration of inflation after 
$\Phi$ passed $\Phi_1$ should secure the number of e-foldings about 30-40 to allow for formation of sufficiently
massive PBH. Though the interaction between $\chi$ and $\Phi$
looks rather artificial, it is not so. This is the general renormalizable 
coupling of two scalar fields.

The second term in eq. (\ref{U-tot}) is the Coleman-Weinberg potential~\cite{CW} 
which arises as a result of one-loop
quantum corrections to $\lambda |\chi|^4 $ interaction.

The remaining two terms are the toy model ones describing flat directions.

The constants $\lambda_1$ and $m$ are generally speaking complex. This may lead to C and CP violation.
However the charge symmetry would be broken only if the relative phase of $\lambda_1$ and $m$ is nonzero.
Coupling of $\chi$ to fermions (quarks) can break C and CP as well.

When $\Phi > \Phi_1$,  potential (\ref{U-tot}) has deep minimum near $\chi =0$ and $\chi$ classically stays
there. In the course of inflation $\Phi$ drops down and at some moment reaches $\Phi_1$. So the barrier
disappears and the window to the flat direction opens. During  this period, when $\Phi$  stays 
close to $\Phi_1$,
the field ${\chi}$ started to  diffuse away from the old minimum,  according to quantum diffusion
equation derived by Starobinsky which we have  generalised to a complex field ${\chi}$. 
At some stage for sufficiently large $\chi$ the diffusion turns into classical motion. 

If the window to the flat direction, when ${\Phi \approx \Phi_1}$ is open only during relatively 
short period, cosmologically small but possibly astronomically large 
bubbles with high $ \beta$ could be created, occupying a small fraction of the universe volume.

Indeed, when $\Phi$ passes the value $\Phi_1$
sufficiently far, the old minimum at $\chi = 0$ reappears and $\chi$ goes back to zero. While $\chi$ is large 
it propagates along the flat direction of the quartic potential. When finally $\chi$ becomes small, it starts to feel
quadratic potential and in the process of motion from quartic to quadratic potentials $\chi$ acquires 
a large angular momentum, that is a large baryonic number, 


If the probability
of $\chi$ to reach a large value is not too big, cosmologically small but possibly astrophysically
large bubbles with high baryon density would be formed
while the rest of the universe has normal
{${{ \beta \approx 6\cdot 10^{-10}}}$, created 
by small ${\chi}$} or by some other mechanism of cosmologicl baryogenesis.

Initially large isocurvature perturbations were generated in chemical content of massless quarks,
while, density perturbations stayed practically zero. 
Density perturbations are generated rather late after the QCD phase transition, when massless
quarks turns into heavy nucleons with masses about 1 GeV, much larger than the temperature 
of the phase transition, $T_{qcd} \sim 100 $ MeV.
The emerging universe looks like a piece of Swiss cheese, where holes are high baryonic 
density objects occupying a minor fraction of the universe volume.  These High-B Bubbles (HBB) 
mostly turn into primordial black holes 


This mechanism of massive PBH formation is quite different from all other known. 
The fundament of PBH creation is build at inflation by making large isocurvature
fluctuations at relatively small scales, with practically vanishing density perturbations, that
appeared only much later. The mass spectrum of PBH reflects the distribution of the bubble by
size during inflation. Its log-normal form is a general feature of diffusion processes.


\section{Summary of the results and conclusion}
$\bullet$ A large lot of outstanding problems of the canonical cosmology can be nicely resolved
if the universe is populated by primordial black holes with masses in the interval from the solar mass up to
supermassive BHs with masses of the order of billion solar masses.\\
$\bullet$ The inverted  mechanism of galaxy formation is proposed when firstly a SMBH was formed and later 
it seeded the galaxy formation by gravitational attraction of the surrounding matter. Thus an existence of SMBH
in almost empty space can be understood.\\
$\bullet$ The early galaxies observed by HST and JWST, when the universe was only several 
hundred  million years old, could be created if seeded by SMBHs.\\
$\bullet$ Existence of a noticeable number of galaxies with masses that are is too small to
allow for creation the observed SMBHs inside, can be explained if these SMBH are primordial.\\
$\bullet$ Creation of SMBH in large contemporary galaxies by conventional accretion mechanism 
demands time larger than the universe age. The problem disappears if the central SPBH is primordial.\\
$\bullet$ Observations of several dwarf galaxies with SMBH in their centres, confirmed our
prediction made several years ago.\\
$\bullet$ The theoretically predicted log-normal mass spectrum of PBH is verified by the chirp mass
distribution of the gravitational waves observed by LIGO/Virgo. The agreement between observation and 
theory is impressively good.\\
$\bullet$  PBHs formed according to our scenario explain the peculiar features of the sources
of GWs observed by LIGO/Virgo, e.g. an existence of  BH with $M= 100 M_\odot$\\
$\bullet$ The density of the intermediate mass black holes (IMBH), $M=(10^2 - 10^5)  M_\odot $
well agrees with their primordial origin. Assumption of the astrophysical formation of IMBH enounters 
serious problems. \\
$\bullet$ Predicted by the model extremely old stars seem to exist even, the existence of the
"older than universe star" can be explained because its old
 age is mimicked by the unusual initial chemistry.  The model also predicts too fast 
 moving stars, which are also observed.\\
$\bullet$ Natural consequence of the suggested model of PBH creation leads to noticeable population of our
Galaxy by antimatter. This striking consequence seems to be confirmed by recent observations.

\section*{Acknowledgement}
The work is supported by  the RSF grant 22-12-00103

\end{document}